\documentclass{article}

\usepackage{xcolor}
\usepackage{PRIMEarxiv}
\usepackage[utf8]{inputenc} 
\usepackage[T1]{fontenc}    
\usepackage{hyperref}       
\usepackage{url}            
\usepackage{booktabs}       
\usepackage{amsfonts}       
\usepackage{nicefrac}       
\usepackage{microtype}      
\usepackage{lipsum}
\usepackage{fancyhdr}       
\usepackage{graphicx}       
\graphicspath{{media/}}     

\title{An entropic analysis of social demonstrations}

\author{
  Daniel Rico \\
  Valencian International University (VIU) \\
  \And
  Yérali Gandica\\
  Valencian International University (VIU) \\
  \texttt{ygandica@gmail.com} \\
}

\begin{document}
\maketitle

\begin{abstract}

Social media has dramatically influenced how individuals and groups express their demands, concerns and aspirations during social demonstrations. The study of X or Twitter hashtags during those events has revealed the presence of some temporal points characterised by high correlation among their participants. It has also been reported that the connectivity presents a modular-to-nested transition at the point of maximum correlation. The present study aims to determine whether it is possible to characterise this transition using entropic-based tools. Our results show that entropic analysis can effectively find the transition point to the nested structure, allowing researchers to know that the transition occurs without the need for a network representation. The entropic analysis also shows that the modular-to-nested transition is characterised not by the diversity in the number of hashtags users post but by how many hashtags they share.

\end{abstract}

keyword:Entropic analysis; social manifestations; nestedness, modular-to-nested transition, complex systems.

\section{Introduction}

Throughout history, social protests have been an effective means for individuals and groups to express their demands, concerns and aspirations in the search for a significant change in several dimensions of society. These collective movements take many forms, from marches to public demonstrations, boycotts and street protests. They all share a common goal: to raise awareness about a specific problem and exert pressure for social change. This constant quest for social transformation has evolved over time, and in the current digital era, social media have emerged as a revolutionary tool that has completely transformed the dynamics of social protests.

The influence of social media in the emergence of demonstrations is undeniable. As stated by Isa et al. \cite{isa2018social}, these digital platforms have provided an instantaneous and global avenue for effervescent individual ideas and opinions to connect and amplify, overcoming geographical and cultural barriers. This phenomenon has radically changed how social protests are organised, promoted, and achieved success in contemporary society. In this context, understanding how social networks influence social demonstrations has become imperative, as this allows us to unravel both the complex interactions and the transfer of mass effect from virtual societies to tangible reality.

Previous studies have pointed to the interplay between online platforms and the subsequent effect of street demonstrations. In a survey of more than 3000 individuals conducted by Gray-Hawkins et al. \cite{gray2018collective}, it was evidenced that 44\% of the interviewees publicly expressed their support for political campaigns on social networks and that $30\%$ attended political demonstrations in the last five years. Similarly, it was observed that $17\%$ of men and $15\%$ of women changed their choices on political or social issues due to information they consumed through social networks. Remarkably, the authors reported that among those who changed their opinion due to social networks, $69\%$ consider that social networks allow them to find people sharing similar opinions on important issues, $65\%$ consider that they allow them to get involved in political and social issues, and $58\%$ consider that social media allow them to express their opinion on such issues. In addition, it was also evident that $14\%$ of the total number of respondents had participated in protests and demonstrations.

Among the various social networks, X (former Twitter) stands out as one of the most impactful platforms for driving social movements. As a public platform, it facilitates the rapid dissemination of information, enabling activists to swiftly make decisions that can trigger prompt mobilisation among its users \cite{conway2015rise,hermida2014sourcing}. X also has hashtags, denoted by the symbol "\#" before a word or phrase. Initially conceived as "channel tags" to allow users to participate in specific discussions, the hashtag has gained recognition for its pivotal role in promoting social movements  \cite{wang2016hashtags}. In \cite{ma2013predicting}, the authors have reported that hashtags work as thematic markers that highlight the relevance of well-known topics and facilitate the effective dissemination of information beyond an individual's network of followers. Using hashtags increases the visibility of a message, as tweets with hashtags are easier to find than text messages. This visibility is crucial to gaining symbolic influence, as it contributes to a rapid and wide dissemination of information. Therefore, the strategic use of hashtags allows the spreading of the content of a tweet to a broader and more diverse audience \cite{wang2016hashtags}. 

Through the tweets' hashtags on four nationwide social demonstrations, in a previous work, Beiro et al. \cite{Beiro2024} have proposed that the high connectivity at specific points of a social demonstration has some characteristics similar to the critical transitions studied in physics. Specifically, it resembles the divergence of the correlation length in those types of transitions. They allege that the simplifications caused in the statistics of the manifestations studied in that article result from those high correlations, as happens in the mentioned critical transitions.

For example, they found that among all demonstrations, the distribution of hashtag frequency shows the highest heterogeneity in the time window during the protests. They argued that this was not due to increased activity but to heterogeneity in user activity. However, there are at least three possible sources for such heterogeneity; one is due to the temporal activity of users. The second one is due to the heterogeneity in how users choose which hashtags to post in their tweets. The last option is related to how users share the hashtags they post. The most relevant heterogeneity type has essential implications for the field of complex systems, as it can shed light on the interplay between the temporal and spatial domains {related to the high correlation occurring at several massive social events}.  

Furthermore, in the same work by Beiro et al., the authors reported that social demonstrations are also characterised by several temporal points of a sustained correlation where hashtags are mostly connected into modular structures. In addition, all the demonstrations presented a transition between that modular interconnectivity and a state characterised by a nested hierarchical structure. Hence, each system passes through several phases characterised by the coordination within subgroups and one state where the system is self-organised into nestedness. A network is said to be perfectly nested when the contacts of a node of a given degree are a subset of the contacts of all the nodes of a higher degree. In terms of the networks of hashtags connected by users posting, this situation means that the hashtags posted by $n$ users are a subset of the hashtags posted by $n + 1$ users. Although the hashtag networks were not perfectly nested, the metrics were high enough to confirm that nestedness was present at those temporal points. \cite{Beiro2024}.  

The transitions modular-to-nested structures have already been studied in social demonstrations \cite{BorgeHolthoefer2017,Beiro2024}. However, the reason for such a dramatic change is still unknown. {Our work is an attempt to understand this phenomenon better. Specifically, we present} an entropic-based study addressing two different points. First, as said before, we are interested in identifying which diversity pattern is the main responsible for the high heterogeneity found in the hashtag frequency distribution reported in \cite{Beiro2024}. Secondly, we also want to answer the question: Is entropy able to find the transition point without the need to build network representations? Payrató-Borràs et al. \cite{PayratoBorras2019} have already conducted an entropic study of nestedness in ecological systems. However, as in most scientific literature, the study of nestedness refers only to static structures that have always been in that state, whereas here, we are interested in using entropic metrics to analyse dynamic structural change during the modular-to-nested transition.

The manuscript continues with a brief description of the events around the three nationwide social demonstrations analysed, and an explanation of the data collection. Then, we explain the construction of the networks. We continue with the definitions used for the different calculus of the entropy-based metrics computed. We also briefly explain the metrics used to compute modularity and nestedness. Next, we show our results, continue with the discussion, and present our overall conclusions in the last section. 
\section{Data and Methods}

\subsection{The historical context of the nation-wide events analysed}

The first dataset (9n) involves a protest against the government's proposed justice reform plans in Argentina on November 9 2019, known as the "9ngranchaporlajusticia". That major demonstration attracted the attention and participation of a wide range of people throughout the country, including opposition groups, civil society organisations and concerned citizens. Using the hashtags "9ngranmarchaporlajusticia" and "9n" in social networks allowed protesters to organise and express their grievances, increasing the reach and influence of the demonstration. This event demonstrated the importance of active citizen participation and civic engagement in Argentina's democratic processes. 

The second dataset involves the "noaltarifazo" protest that also occurred in Argentina between 4 and 6 January of the same year. During this protest, citizens expressed their discontent with government policies, particularly those related to taxes and the cost of public essential services such as electricity and gas. Using the hashtags "noaltarifazo" and "ruidazonacional" in social networks played a crucial role in mobilising and organising protesters, allowing them to coordinate their efforts and share information about the demonstration. Furthermore, it demonstrated the power of social media to facilitate public discontent with government policies.

Finally, the last dataset is related to the tragic event that marked France in January 2015. The terrorist attack on the offices of the satirical magazine Charlie Hebdo in Paris shocked the country. The assailants targeted the magazine for publishing controversial caricatures of the Prophet Muhammad. The attack resulted in the death of twelve people, including prominent cartoonists, and sparked debates worldwide on freedom of expression, extremism, and national security. As is customary, these responses swiftly inundated social media platforms, with Twitter being the focal point. Commencing on January 7, millions of tweets surfaced employing hashtags like \#CharlieHebdo and \#JesuisCharlie,    \cite{enwiki:1182928449} \cite{enwiki:1188883112}\cite{smyrnaios2017charlie} which resulted in a massive demonstration on January, the 11th.

\subsection{Data collection}

We found the two most used hashtags for each protest during the event day(s), as already mentioned for each demonstration. Next, we created the universe of users, listing all users who tweeted at least one of those two hashtags on the event day(s), shown in table ~\ref{Datasets}. Finally, we collected all the hashtags posted by the universe of users over a broader period, which is also displayed in the same table. 

\begin{table}
    \centering
    \begin{tabular}{|c|c|c|c|} \hline 
         Dataset&Event day(s) & Data collection & N. Hashtags\\ \hline 
         9n& 9 Nov 2019 & 8/11, 6am - 10/11, 8pm & 18.193\\ \hline 
         Noaltarifazo& 4-6 Jan 2019 & 1/1, 6 am - 7/1, 11pm & 22.813\\ \hline
          CharlieHebdo & 11 Jan 2015 & 10/1, 1am - 12/1, 23 pm & 17.638
          \\\hline
    \end{tabular}
\caption{Composition of the data sets. } 
\label{Datasets}
\end{table}

\subsection{Construction of the networks}
Our one-hour temporal networks have been built so that nodes are hashtags. The link weight between two hashtags represents the number of users who posted that pair of hashtags during that hour. We compute modularity and nestedness on our temporal networks. Modularity quantifies the presence of community structure in the network, and we use the Louvain method \cite{Blondel2008} from the NetworkX package. In general terms, modularity compares the number of edges observed within clusters and what is expected in a comparable-size network in which edges or links are randomly distributed. High modularity means dense intra-community connections but sparse inter-community ones.  

We also quantify the presence of nested structure in our networks, i.e., the neighbourhood of each node is contained in the neighbourhood of nodes with higher degrees. For measuring nestedness, we used the Nestedness Calculator for Python based on the measurement proposed in \cite{Almeida‐Neto2008}. Both self-organisations --modular and nested-- are incompatible as the first one is arranged into low-connected communities while a hierarchical structure characterises the last one.  

\subsection{Entropy}

We calculate the entropy as defined in information theory \cite{Cover2006}, namely as the uncertainty or variability of the probabilities of a specific output. We calculate the entropy or variability in each hour, $h$, by:

$$H(X)^h=- \sum_{x \in X} p(x) log p(x).$$

We compute the entropy for four different ways of defining the probabilities. First, we calculate the variability of hashtags per hour. For this purpose, we compute, for each hour, the frequency of each hashtag existing in that hour. In this way, we obtain the probability that a user, from the universe of users in that hour, posts each of the Twitter hashtags used in that hour. In that sense, a high entropy value refers to the situation where some hashtags are posted very little while others are posted quite a lot. Conversely, a low entropy value is related to a situation where either only one (or very few) hashtag(s) is (are) posted or all of them are posted with the same frequency.

Secondly, we perform the same calculation for users. After computing the frequency of each user, i.e., the number of hashtags (whether repeated or not) posted in each hour, we calculate the entropy. Hence, low entropy means that only one (or very few) user(s) posts in that hour or that everyone posts the same number of times. Maximum entropy is the case where users posting few and many hashtags are equally likely.

We are interested in the diversity connecting users and hashtags in what follows. First, we compute the diversity of users per hashtag. For this purpose, we group by hashtags in each hour and compute the frequency of users (normalised number of unique users). This probability in the entropy calculation allows us to quantify the variability in how hashtags are more or less shared by different users in that hour. A low entropy value means that roughly all hashtags have been posted by the same number of users. On the other hand, a high entropy indicates that some hashtags were posted by only a few users while others were highly posted.

Finally, we are also interested in quantifying the variability of how users post different hashtags each hour, i.e., the variability of hashtags per user. It makes a difference whether a user posts the same hashtag (or a few of them) or different ones during the hour $h$. For this endeavour, we grouped by users and computed the normalised number of unique hashtags in each hour. Henceforth, high entropy means little difference between the number of users posting a few and many different hashtags.  

\subsection{Modularity and nestedness}

We calculate the modularity in the networks using the community module of the Python 'python-louvain' library. The equation for modularity in this library is based on the original definition proposed by Newman and Girvan \cite{girvan2002community}: 
$$Q = \frac{1}{2m} \sum_{ij} (A_{ij} - \frac{k_{i}k_{j}}{2m}) \delta(c_i, c_j)$$
Where:
    \begin{itemize}
        \item $Q$ is the modularity score, $m$ is the total number of edges in the network.
        \item  $A_{ij}$ is the element in the adjacency matrix representing the connection between nodes $i$ and $j$
        \item  $k_{i}$$k_{j}$are the degrees of nodes $i$ and $j$, respectively, representing the number of edges connected to each node.
        \item $c_i$, and  $c_j$ are the community assignments of nodes  $i$ and $j$, respectively.
        \item   $\delta(c_i, c_j)$ is the Kronecker delta function, which is 1 nodes $i$ and $j$ are in the same community and 0 otherwise.
    \end{itemize}

Similarly,  we calculate the nestedness in the networks using the NODF (Nestedness metric based on Overlap and Decreasing Fill) index \cite{almeida2008consistent}:
$${NODF} = \frac{\sum_{i,j} \min(O_{ij}, O_{ji})}{\sum_{i,j} O_{ij}}$$
The NODF index compares the overlap ($O$) of nodes between modules in a network to the total overlap expected in a perfectly nested network. 
 
\section{Results}

\subsection{Modularity and Nestedness}

\begin{figure}[h]
\includegraphics[width=10.5 cm]{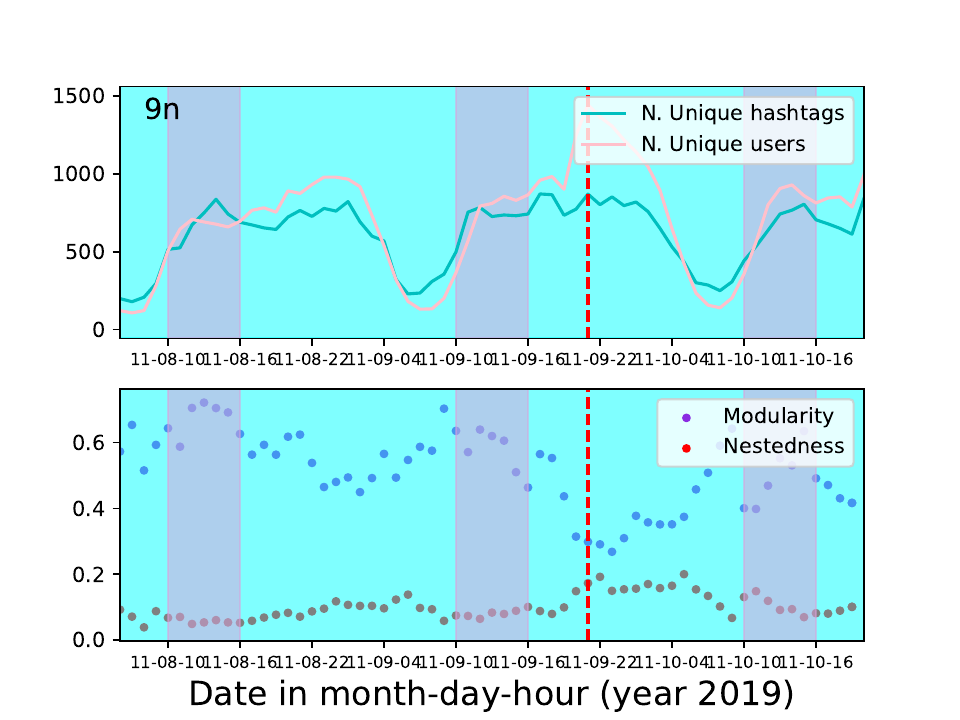} 
\hspace{-2.3cm} \\ 
\includegraphics[width=10.5 cm] {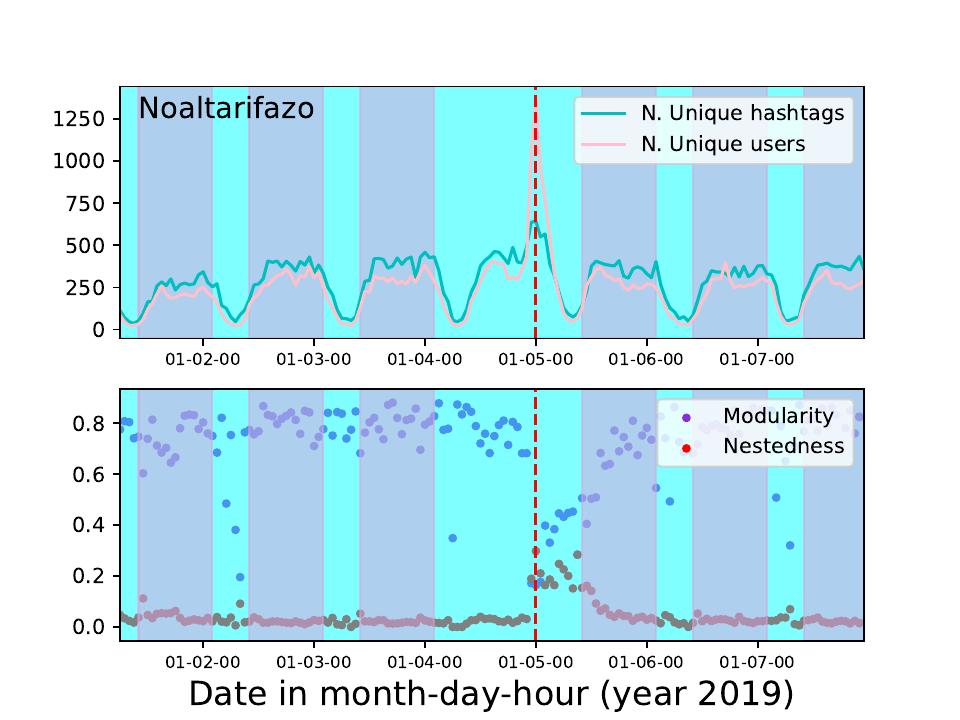} 
\hspace{-2.3cm}
\includegraphics[width=10.5 cm] {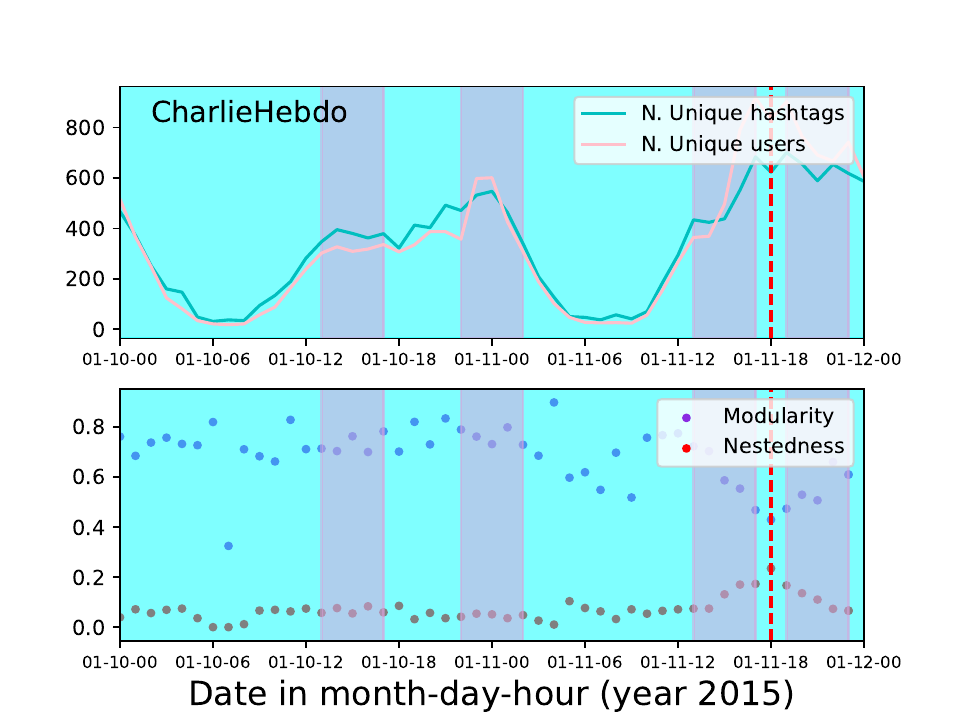} 
\caption{The number of unique hashtags and unique users in each one-hour temporal point are shown in the upper part. The bottom part depicts the values for modularity and nestedness on each one-hour temporal network. The background highlights in violet the temporal points that meet two conditions: high modularity and excluding inactive hours. In blue is signalled the points of low activity. A dashed-red line shows the point of highest nestedness. Results are shown for the demonstrations 9n, Noaltarifazo and CharlieHebdo.
\label{fig4}}
\end{figure}

In figure \ref{fig4}, we show, in the upper panel, the number of unique hashtags and unique users in each one-hour temporal point. In the bottom part of the same figure, we show the calculations for modularity and nestedness for the same one-hour temporal networks. We can see in that plot the points of high modularity and the quasi-instantaneous transition from modular to nested structure (signalled by dashed-red lines). {Notice there is no mathematical definition for that transition; the transition is structural on the network representations.}  

We used these results to construct the two-colour background adopted in that figure and the subsequent ones throughout the paper. The background highlights in violet the temporal points that meet two conditions: high modularity and excluding inactivity. In blue is signalled the points of low activity, presumably because people are mostly sleeping. A dashed-red line shows the point of highest nestedness. That region was left in blue because the modularity decreases immediately after the modular-to-nested transition. In each demonstration, the violet regions have all the same width, showing the cyclical nature of those temporal windows, which were associated with a sustained correlation in \cite{Beiro2024}.  
\hspace{-3cm}

\subsection{Entropic analysis}

\begin{figure}[h]
\hspace{-4 cm}
\includegraphics[width=18 cm]{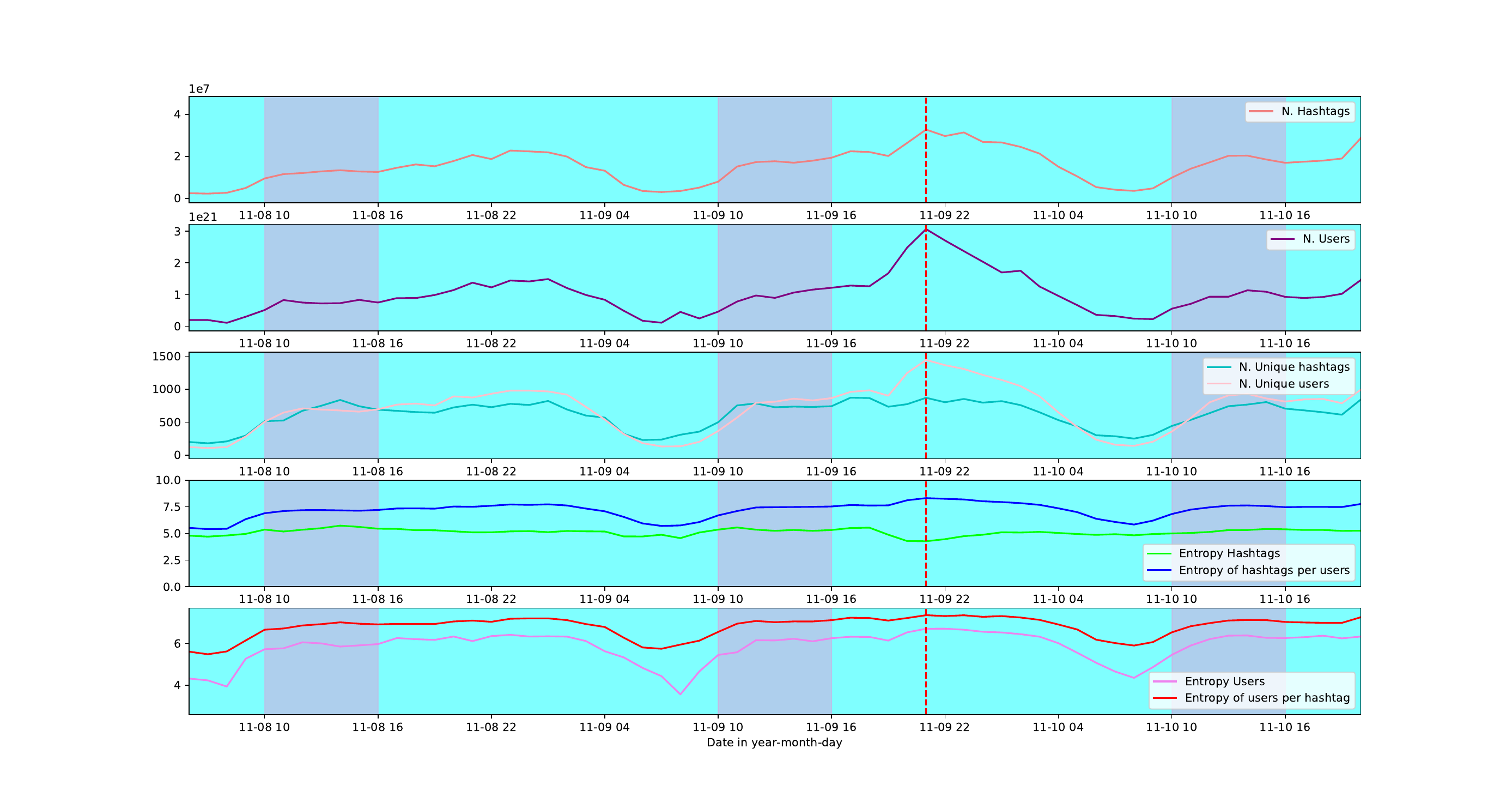} 
\caption{It is shown, from top to bottom, the number of hashtags, the number of users, the unique value of both numbers, the entropy for users along with the entropy of users per hashtag, and the entropy of hashtags along with the entropy of hashtags per user. All the plots correspond to the social demonstration 9n. With two different colours, the background highlights the temporal points of high modularity (violet), and a dashed-red line shows the point of highest nestedness. \label{fig1}}
\end{figure}

In figure \ref{fig1}, we show for the social demonstration 9n, from top to bottom, the number of hashtags, the number of users, the unique value of both numbers, the entropy of hashtags along with the entropy of hashtags per user, and finally, the entropy for users along with the entropy of users per hashtag. The first observation is that the number of unique users and unique hashtags are generally of the same order. However, we can see that this scenario changes at the point of highest activity, corresponding to the point of maximum nestedness (red line). At that point, the number of unique users suddenly increases while the number of unique hashtags decreases. Consequently, that temporal point attains the maximum distance between both values. 

The fifth panel of the same figure \ref{fig1} also shows that both the user's entropy and the users' entropy per hashtags are high at the critical point; nevertheless, they are still in the same range as at points of high modularity (violet zones). However, this is not the case for hashtags (fourth panel). The most exciting sign shown by the figure is the change in the hashtag's entropy and the entropy of hashtags per user. There is a decrease in the diversity of hashtags at that point. However, at the same time, there is maximum heterogeneity in the way how hashtags are more or less shared by different users in that hour. That fact seems to be the main characteristic of the system in the nested configuration.  

\begin{figure}[h]
\hspace{-4 cm}
\includegraphics[width=18 cm]{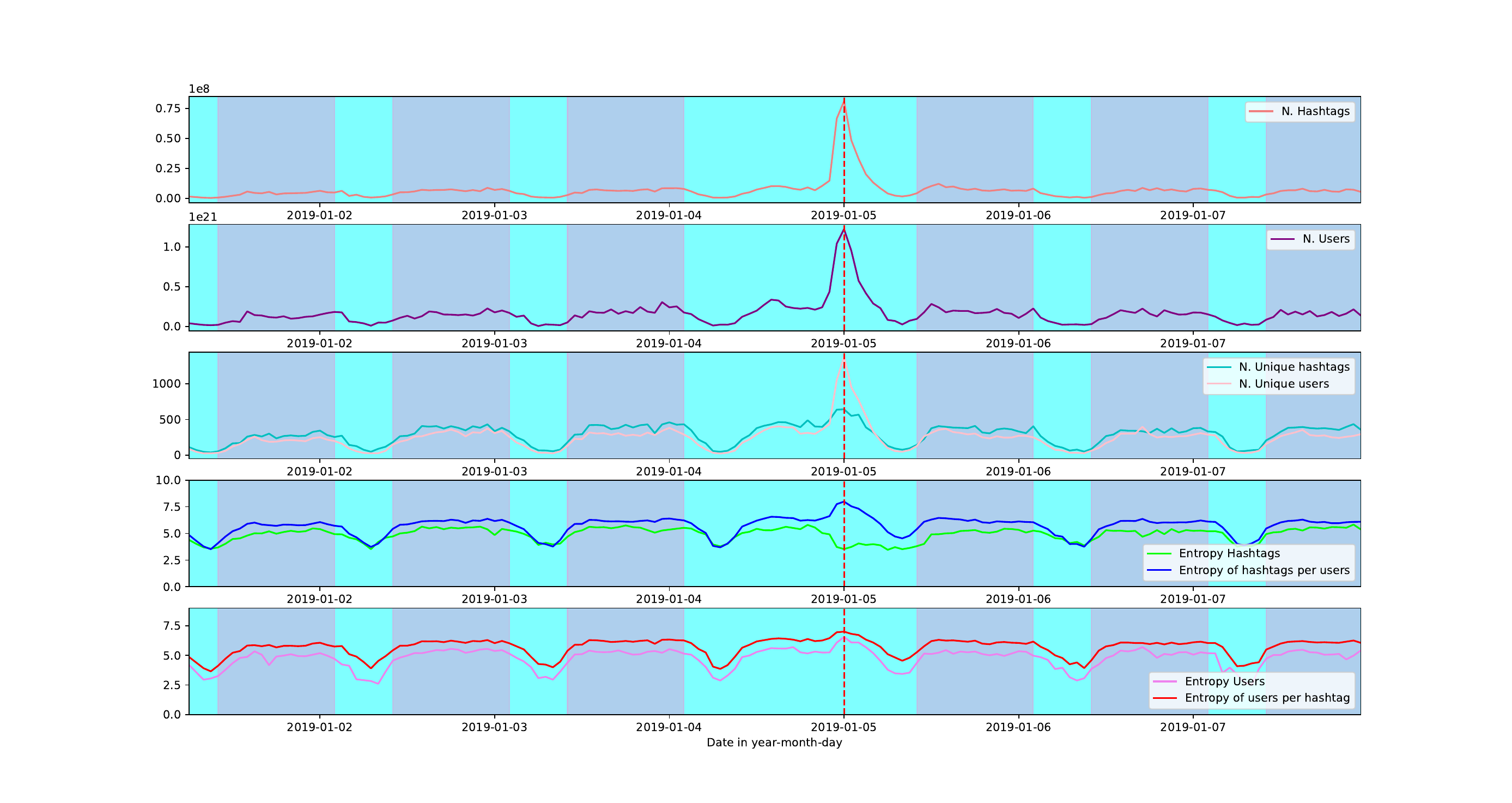}
\caption{It is shown from top to bottom the number of hashtags, the number of users, the unique value of both numbers, the entropy for users along with the entropy of users per hashtag, and the entropy of hashtags along with the entropy of hashtags per user. All the plots correspond to the social demonstration Noaltarifazo. With two different colours, the background highlights the temporal points of high modularity (violet), and a dashed-red line shows the point of highest nestedness. \label{fig2}}
\end{figure}

In figure \ref{fig2}, the same scheme is shown but for the demonstration Noaltarifazo. In this case, we can see that the number of unique users and hashtags are closer. The same happens for their entropy during the active hours, with the only difference being at the point of nested behaviour. As in the case of the 9n demonstration, the critical point is characterised by the lowest hashtag entropy (during active hours) while the highest entropy of hashtags per user.

\begin{figure}[h]
\hspace{-4 cm}
\includegraphics[width=18 cm]{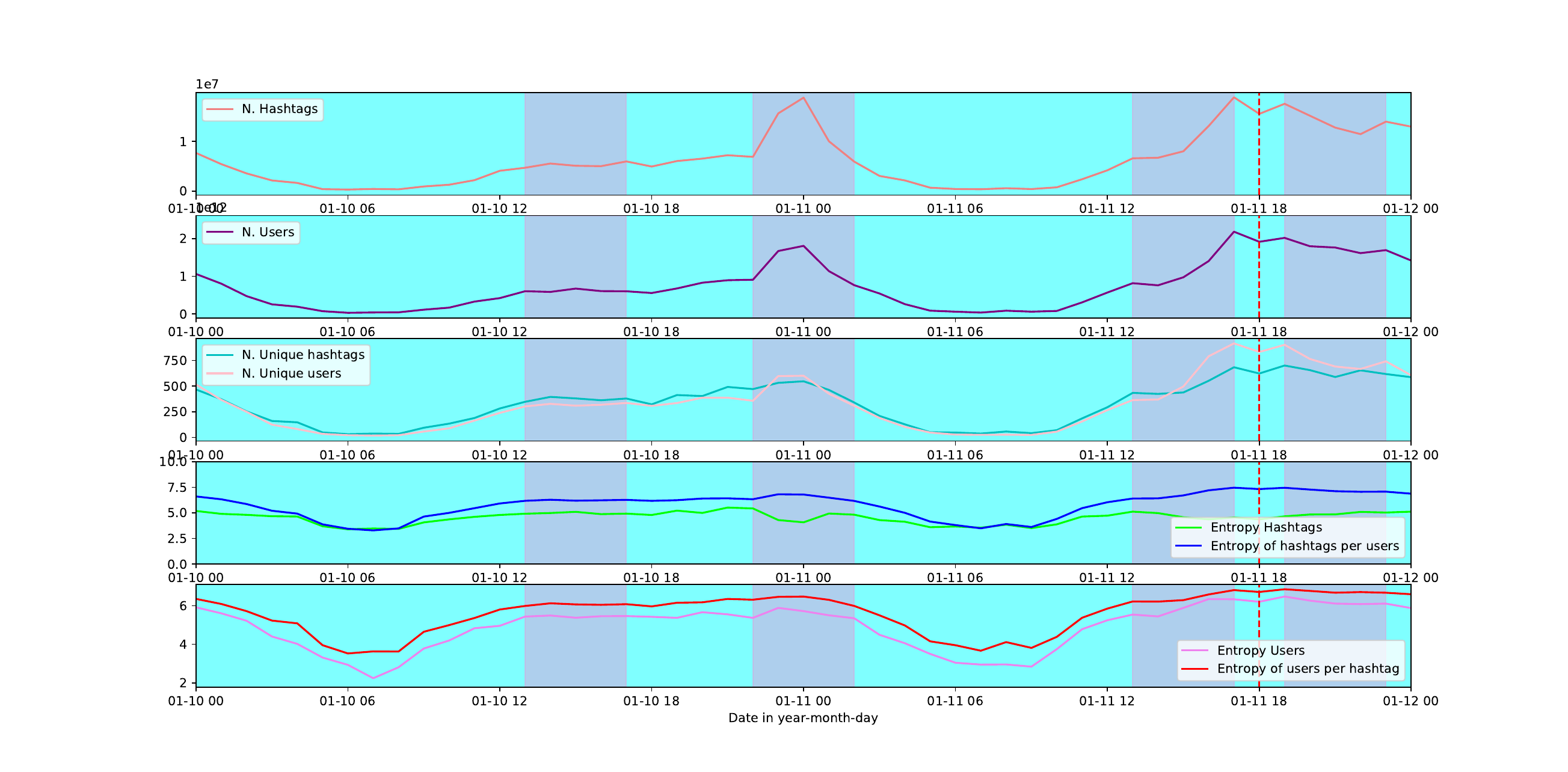}
\caption{It is shown from top to bottom the number of hashtags, the number of users, the unique value of both numbers, the entropy for users along with the entropy of users per hashtag, and the entropy of hashtags along with the entropy of hashtags per user. All the plots correspond to the social demonstration CharlieHebdo. With two different colours, the background highlights the temporal points of high modularity (violet), and a dashed-red line shows the point of highest nestedness. \label{fig3}}
\end{figure}

The demonstration Charlie Hebdo confirms the last results. As we can see in figure \ref{fig3}, even though the users' entropy and users' entropy per hashtag are high, the characteristics that determine the critical point (red line) is the lowest value of the hashtag entropy while the highest one on the entropy of hashtags per user. 

\section{Discussion}

Our results are consistent over the three demonstrations analysed. First, looking at the points of active hours, we can see that, in general, the four entropic-based measurements remain at stable values except at the point of the highest activity, signalled by the red line. At that point, there is a decrease (increase) in the entropy of hashtags (per user). The last can be understood by the nested structure of the system at that critical point. Namely, hashtags posted by $N$ users, with considerable statistics, are a subset of the hashtags posted by $N+1$ users. This will inevitably lead to a configuration characterised by less diversity of hashtags but, at the same time, high variability in how users share them due to the hierarchical structure in the hashtags posting.

Regarding the users' entropy and entropy of users per hashtag, we can not say that they present a significant behaviour at the critical point (red line). However, they both attain their maximum values at that temporal point. With respect to our first research question about the primary source of heterogeneity in user activity, we can say that it is how users share hashtags that is more variable. Finally, regarding the second research question, entropy-based measures can well characterise when a system presents nestedness.

{Concerning the possible limitations of our study, we could say that the entropic study does not depend on the network representation, which eliminates possible sources of bias related to the type of representation. However, the data itself could suffer from bias. For example, other points of high nestedness or modularity could exist, not captured by the Twitter data but by other media, such as other social networks. The last is a difficult limitation to overcome, as studies are often based on a single data source.}  

\section{Conclusions}

The results obtained in this work show that an entropic approach can detect nested configurations. This result is interesting because nested structures result from connectivity, and the entropic approach can be used without needing a network representation. Researchers can apply this methodology to the data before figuring out how to define the network topology. The same result also allows us to conclude that the fact that we only found such a characterisation at the point indicated by the red line leads us to infer that there is no hierarchical structure within the communities during the modular phases, which could have been a plausible hypothesis \cite{Palazzi2021}.

Finally, we would like to highlight that the time points presenting nestedness in massive demonstrations are characterised by a high correlation between their participants. Those points were associated with critical social situations in which the action of a single individual can reach a national dimension. In that sense, it is crucial to find more systems exhibiting a transition to the hierarchical nested structure in order to better understand their emergence.\\

\section*{Authors contributions}
Conceptualization, Y.G.; methodology, Y.G.; software, D.R. and Y.G.; validation, Y.G.; formal analysis, Y.G.;  resources, Y.G.; data curation, Y.G.; writing D.R, and Y.G; supervision, Y.G.; All authors have read and agreed to the published version of the manuscript.

\section*{Funding}
The data collection was funded by Labex MME-DII (Grant No. ANR reference 11-LABX-0023-01) granted to Y.G.'' The data was collected by the company DATA FOR SCIENCE, INC.

\section*{Data availability}
The tweet ids' datasets generated for the two Argentinian protests during the current study are available in the repository: \url{https://github.com/yerali/Ids_for_two_Argentine_social_movements}

\bibliographystyle{spphys}
\bibliography{references}
\end{document}